\documentclass[12pt, draftclsnofoot, onecolumn]{IEEEtran}

\usepackage{algorithm}
\usepackage{algorithmic}
\usepackage{cite}
\usepackage{graphicx}
\usepackage{amsmath}
\usepackage{amssymb}    
\usepackage{amsfonts}
\usepackage{array,booktabs}
\usepackage[flushleft]{threeparttable}
\usepackage{times}
\usepackage{epsfig}
\usepackage[latin5]{inputenc}
\usepackage{setspace}
\usepackage{booktabs}
\usepackage[mathscr]{euscript}
\usepackage{color}
\usepackage{array,multirow}
\usepackage{epstopdf}
\usepackage{pifont}
\usepackage{multirow}
\usepackage{hhline}
\usepackage{slashbox}
\usepackage{smartdiagram}
\usepackage{caption}
\usepackage{subcaption}
\usepackage{float}
\usepackage[most]{tcolorbox}
\usepackage{colortbl}
\usepackage{makecell}

\definecolor{lgray}{gray}{0.9}
\colorlet{dgray}{lightgray!40}

\colorlet{MyBlue}{red!25!green!50!blue!50}
\colorlet{lblue}{MyBlue!10}
\colorlet{mblue}{MyBlue!20}
\colorlet{dblue}{MyBlue!30}

\tikzset{
  basic/.style  = {draw, text width=2cm, drop shadow, font=\sffamily, rectangle},
  root/.style   = {basic, rounded corners=2pt, thin, align=center,
                   fill=green!30},
  level 2/.style = {basic, rounded corners=6pt, thin,align=center, fill=green!60,
                   text width=8em},
  level 3/.style = {basic, thin, align=left, fill=pink!60, text width=6.5em}
}

\newcolumntype{L}[1]{>{\raggedright\let\newline\\\arraybackslash\hspace{0pt}}m{#1}}
\newcolumntype{C}[1]{>{\centering\let\newline\\\arraybackslash\hspace{0pt}}m{#1}}
\newcolumntype{R}[1]{>{\raggedleft\let\newline\\\arraybackslash\hspace{0pt}}m{#1}}

\begin{document}

\title{Emerging Edge Computing Technologies for Distributed Internet of Things (IoT) Systems}

\author{\IEEEauthorblockN{Ali~Alnoman, Shree~Krishna Sharma, Waleed~Ejaz, and Alagan~Anpalagan}
\thanks{A. Alnoman and A. Anpalgan are with the Department of Electrical and Computer Engineering, Ryerson University, 350 Victoria St., Toronto, Canada, S. K. Sharma is with the SnT, University of Luxembourg, Luxembourg, W. Ejaz is with Thomson River University, British Columbia, Canada.}
}

\markboth{IEEE Wireless Communications Magazine (Draft)}{}

\maketitle

\begin{abstract}
The ever-increasing growth in the number of connected smart devices and various Internet of Things (IoT) verticals is leading to a crucial challenge of handling massive amount of raw data generated from distributed IoT systems and providing real-time feedback to the end-users. Although existing cloud-computing paradigm has an enormous amount of virtual computing power and storage capacity, it is not suitable for latency-sensitive applications and distributed systems due to the involved latency and its centralized mode of operation. To this end, edge/fog computing has recently emerged as the next generation of computing systems for extending cloud-computing functions to the edges of the network. Despite several benefits of edge computing such as geo-distribution, mobility support and location awareness, various communication and computing related challenges need to be addressed in realizing edge computing technologies for future IoT systems. In this regard, this paper provides a holistic view on the current issues and effective solutions by classifying the emerging technologies in regard to the joint coordination of radio and computing resources, system optimization and intelligent resource management. Furthermore, an optimization framework for edge-IoT systems is proposed to enhance various performance metrics such as throughput, delay, resource utilization and energy consumption. Finally, a Machine Learning (ML) based case study is presented along with some numerical results to illustrate the significance of edge computing.
\end{abstract}


\section{Introduction}
Bringing smart sensors and Machine-Type Communications (MTC) devices around us to facilitate our daily life will characterize the next era of information and communication technology. From smart homes and transportation to the industrial manufacturing and supply chain, Internet of Things (IoT) is prevailing the attention of industries and academia nowadays. The number of connected devices is forecasted to reach around $125$ billion (IHS Markit) by 2030, and the amount of data generated by the connected devices is estimated to be about $507$ zettabytes per year (Cisco) by 2019 \cite{sun2016}. This remarkable increase in demand for resources must be accompanied by a similar increase in resource provisioning to avoid any form of resource deficits or service outages.

The massive amount of data originated from distributed IoT devices can be well handled by the existing cloud computing platform due to its huge computational and storage capacity. However, because of its centralized mode of operation and the involved delay, this is not suitable for distributed IoT systems and for the real-time operation of latency-sensitive IoT applications. To this end, providing cloud-like computing, storage and communication facilities at the network edge helps to reduce the end-to-end latency, to save bandwidth resources in the backhaul links, and also to alleviate the computational burdens on the cloud-servers. In this direction, the European Telecommunications Standards Institute (ETSI) has already standardized and defined Mobile Edge Computing (MEC) as the provision of computing services within the Radio Access Network (RAN) in close proximity to mobile connected devices \cite{abbas2017}.

Fostering the cooperation between cloud and edge computing is essential to ease the processing and analysis of a huge amount of unstructured IoT data. In addition to providing computing capabilities at the vicinity of users, edge computing devices can perform various preliminary tasks including data classification and filtering, service-level agreement ranking and parameter measurements before involving the central-cloud. Also, collaborative processing between edge and cloud computing with regard to resource allocation, load balancing and task offloading can provide a strong support to both delay-sensitive and computing-intensive tasks in wireless IoT networks \cite{Sharma2017live}.



Not only computing resources that recall for effective solutions, but also the radio resources which enable the connectivity of IoT devices. Herein, one of the envisioned solutions to cope with the scarcity of radio resources is to integrate all available Radio Access Technologies (RATs) such as 5G, LTE, WiFi and fiber optics. In addition, the deployment of a large number of low-power small Base Stations (BSs) can enhance the cellular network capacity by exploiting the spatial frequency reuse over small geographical areas. Nevertheless, systems with such radio and computing heterogeneity require sophisticated management and control schemes, which can be enabled with the recent advances in Network Function Virtualization (NFV) and Software-Defined Networking (SDN) technologies. In this regard, Heterogeneous Cloud RANs (H-CRANs) wherein all network nodes enter a shared processing entity, namely Baseband Unit (BBU) pool, is a practical implementation scenario for NFV and SDN technologies towards providing effective management of resources and infrastructure.

Most existing research works have focused on the computing aspects of IoT systems without providing connections to the emerging cellular advances especially those related to H-CRANs, or how to establish a high-level coordination between cellular and computing infrastructures. Moreover, SDN, big data analytics, and artificial intelligence are usually introduced as application-specific technologies rather than being adopted within a fundamental optimization framework. From the aforementioned, we aim to present comprehensive insights on distributed IoT systems taking into account of both the communication and computing challenges and technologies fostering adaptive, elastic, and learning-based solutions. The main contributions of this paper are highlighted below:
\begin{itemize}
  \item Introduce the main practical challenges facing edge-IoT computing and propose potential solutions.
  \item Provide a classification of the emerging technologies in edge-IoT systems into different sub-categories.
  \item Propose an optimization framework to tackle different system-level aspects such as computing, delay, scheduling, and energy consumption.
  \item Present a Machine Learning (ML)-based case study regarding edge-computing for a distributed IoT system.
\end{itemize}

\section{Edge Computing for Distributed IoT Systems: Challenges and Potential Solutions}
The heterogeneous nature of distributed IoT systems regarding devices, applications and service provisioning poses additional technical challenges for the effective management of computing and communication resources. Table \ref{challenges} summarizes the major challenges and provides potential enabling solutions for edge-IoT systems under three main domains, which are briefly described below.
{\fontsize{11pt}{12.0pt}\setlength{\arrayrulewidth}{0.4mm}
\begin{table*}[h]
  \caption{\small{Challenges and Potential Solutions for Edge-IoT Systems}}\label{challenges}
\begin{center}
\begin{tabular}{p{3cm}|p{5cm}|p{7cm}}
\hline
\cellcolor{dblue} \textbf{Main domains} & \cellcolor{mblue} \textbf{Challenges} &  \cellcolor{dblue} \textbf{Potential solutions} \\ \hline \hline

\multirow{4}{*}{\cellcolor{dblue}} & \cellcolor{mblue} Delay & \cellcolor{dblue} Cooperative cloud-edge computing and intelligent task offloading mechanisms \\
\hhline{>{\arrayrulecolor{dblue}}->{\arrayrulecolor{black}}|--}

\cellcolor{dblue} \makecell{Diversity in IoT} & \cellcolor{mblue} Mobility & \cellcolor{dblue} BBU pool, distributed fog nodes, and Ad-hoc fogs \\
\hhline{>{\arrayrulecolor{dblue}}->{\arrayrulecolor{black}}|--}
\cellcolor{dblue} & \cellcolor{mblue} Security and privacy & \cellcolor{dblue} Authentication protocols \\ \hline

\multirow{4}{*}{\cellcolor{dblue}}   & \cellcolor{mblue} Insufficient computing resources & \cellcolor{dblue} Bi-directional resource sharing between edge (fog) and central (cloud) servers \\  \hhline{>{\arrayrulecolor{dblue}}->{\arrayrulecolor{black}}|--}

\cellcolor{dblue} \makecell{Resource \\management} & \cellcolor{mblue} High demand on a particular content & \cellcolor{dblue} In-network caching \\  \hhline{>{\arrayrulecolor{dblue}}->{\arrayrulecolor{black}}|--}

\cellcolor{dblue} & \cellcolor{mblue} Redundant data transmission & \cellcolor{dblue} Data aggregation and analysis \\ \hhline{>{\arrayrulecolor{dblue}}->{\arrayrulecolor{black}}|--}
\cellcolor{dblue}  & \cellcolor{mblue} More demand on the uplink & \cellcolor{dblue} Scheduling techniques (e.g., TDMA) \\ \hline

\multirow{3}{*}{\cellcolor{dblue}}  & \cellcolor{mblue} Various computing entities & \cellcolor{dblue} Hierarchical architectures (e.g., fog-to-cloud and cloud-to-fog) \\  \hhline{>{\arrayrulecolor{dblue}}->{\arrayrulecolor{black}}|--}

\cellcolor{dblue} \makecell{Performance \\ coordination} & \cellcolor{mblue} Multiple service providers &  \cellcolor{dblue} Standardization and utilization of compatible infrastructures and platforms \\  \hhline{>{\arrayrulecolor{dblue}}->{\arrayrulecolor{black}}|--}
\cellcolor{dblue}  & \cellcolor{mblue} Interoperability & \cellcolor{dblue} SDN and NFV \\ \hline
     \end{tabular}
  \end{center}
\end{table*}
}
\subsection{Diversity of IoT Requirements}
The deployment of IoT applications in a wide variety of our daily life applications requires distinct Quality of Experience (QoE) provisioning for each particular application. For instance, delay tolerance in autonomous driving and biomedical sensors should not exceed a few milliseconds, whereas in other applications such as climate monitoring, several minutes of delay can be tolerated. Also, MTC is characterized with bursty and low data-rate transmission. Therefore,  classification of IoT data into different service categories can facilitate in investigating various aspects including QoE requirement, security level, resource demand, and the suitability of cloud or edge computing for a particular application. From the energy-delay tradeoff perspective, it is necessary to coordinate between edge-users and edge-devices during the offloading process to minimize energy consumption. For instance, task offloading in some circumstances can incur higher energy consumption than performing on-device processing due to communication overhead \cite{meng2017}.

\subsection{Resource Management}
The storm of requests from mobile devices can deteriorate, or even crash the core network due to the limited capability of mobility management entity \cite{zhou2014}. Furthermore, according to real-time measurements in a wideband network, about $90$\% cell congestions are separated by less than $13$ minutes, and about $90$\% congestions last for less than $1.2$ seconds \cite{zhou2014}. These challenges cause bottlenecks in the implementation of practical IoT systems, and therefore, hybrid access control schemes are required to combine the merits of both the scheduled and random access schemes based on real-time system parameters  \cite{alexiou2014}. Moreover, there is an indispensable need to devise elastic radio resource allocation schemes that adapt to the temporary demands of edge networks. In the context of H-CRANs, the BBU pool can provide substantial assistance since all network resources are virtualized and managed by the unified servers.

Another main feature of fog computing is volatility which enables system flexibility in responding to dynamic IoT requirements and reduces workload on the computing infrastructure. For example, vehicular clouds can form intermittent fog-clouds in parking lots or even in highways by sharing information, locations, awareness messages while coexisting in a particular area. These information can be exchanged locally among vehicles, and also with smart traffic lights, street lamps and road-side infrastructure.

\subsection{Performance Coordination}
The integration of a broad variety of services, devices, and RANs in one architecture imposes many challenges in terms of Quality of Service (QoS) provisioning, compatibility, load sharing and synchronization. Therefore, distributed network intelligence is essential due to the large-scale and high complexity of IoT systems, and to provide system elements with decision making capabilities towards a smart IoT environment \cite{sahni2017}. For instance, although cloud-servers have more powerful capabilities, the round-trip time between IoT devices and the cloud may not satisfy the desired QoE of delay-sensitive applications. Here, efficient coordination among different fog layers, and also in between fog layers and the central-cloud must be maintained to minimize the latency experienced by the end-users.

Unlike the MEC by ETSI, OpenFog Consortium has not yet standardized the fog computing, especially in regard with interoperability and security among multiple service providers and computing entities \cite{AI2017}. However, communication standardization seems more encouraging where the 3GPP Release 13 revealed the Narrowband IoT technology to enable IoT services using low-power wide area networking.

\section{Classification of Emerging Edge-IoT Technologies}
\label{sec:resource}
\begin{figure*}[h]
\centerline{
\includegraphics[scale=0.55]{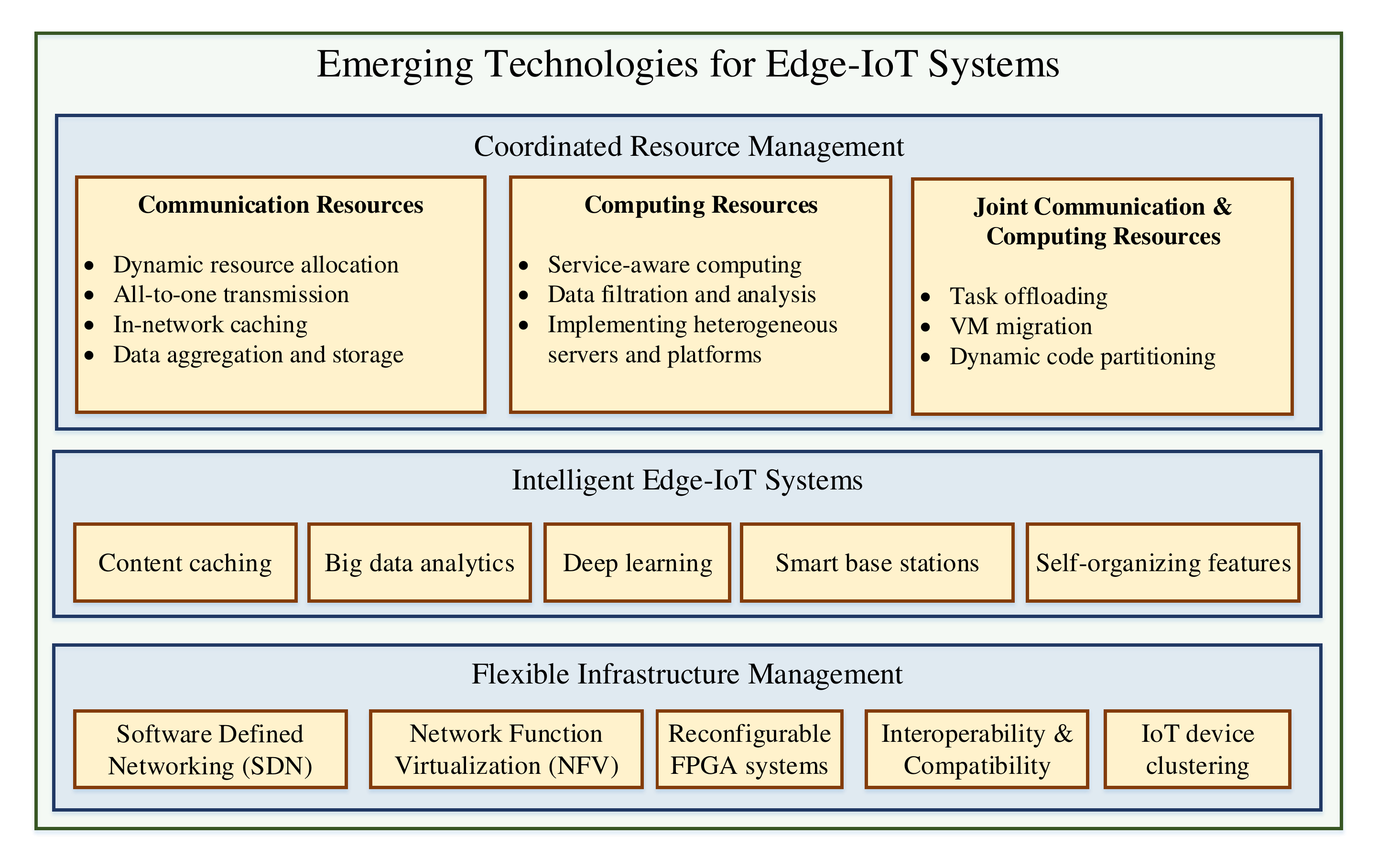}
}
\caption{\small{Classification of emerging edge computing technologies for IoT.}}\label{fig:class}
\end{figure*}

The diverse QoS requirements in IoT systems, which inherently integrate both cellular and computing services, necessitate the employment of advanced communications and computing technologies. Also, data analytics and network intelligence can play a pivotal role in such systems to tackle the unprecedented amounts and the unpredicted behaviour of connected devices. Here, we classify the enabling technologies for edge-IoT systems into three categories as highlighted in Fig. \ref{fig:class} and briefly discuss them below.

\subsection{Coordinated Resource Management}
\subsubsection{Communication Resources}
The ultra-dense deployment of SBSs in future cellular networks requires sophisticated resource management mechanisms to improve both spectral and energy efficiencies, and to reduce the cross-tier interference \cite{wang2017}. Furthermore, the overhead of information exchange among SBSs, Macro BSs (MBSs) and the core network could limit the overall system performance. To this end, Cloud RAN (C-RAN) is considered a promising technology which allows efficient coordination among network elements in a unified controller (BBU pool), and multiple radio access technologies can be integrated to maximize the network capacity via this SDN-based central controller. For instance, 5G, LTE, WiFi, Coordinated Multi-Point (CoMP), millimeter wave (mmWave), massive-MIMO, and Non-Orthogonal Multiple Access (NOMA) technologies can operate concurrently with less information exchange among different network layers and nodes \cite{alexiou2014}. Moreover, instead of offloading all tasks to the edge-servers, Device-to-Device (D2D) technology can reduce the burdens on the servers by enabling mobile users to benefit from the available computing resources in nearby devices in a cooperative manner \cite{sahni2017}.


\subsubsection{Computing Resources}
Towards the adaptive provisioning of computing resources with satisfactory QoE such as delay, energy and Central Processing Unit (CPU) cycles, it is essential to establish a hierarchical architecture between cloud and edge-devices. The Fog-to-Cloud (F2C) paradigm \cite{bruin2016} is an example of the cloud-fog cooperative performance that integrates resources of cloud and fog networks to leverage the overall computing capabilities. In future computing frameworks, every node including user devices could take part in the computing tasks. The concept of consumer as a provider is one of the implementations that allow user devices to share their available computing capabilities with ambient devices, or further, with the fog and cloud-servers. Moreover, vehicular and drone-based fogs facilitate in providing on-demand computing services to adjacent mobile users with low latency.

\subsubsection{Joint Communication and Computing Resources}
Equipping SBSs of future cellular networks with computing functions will turn SBSs into the so-called 'smart SBSs'. In an LTE-based system, small-cell cloud enhanced eNodeB is a practical implementation of smart SBSs whereby cellular, cloud and edge computing function together \cite{barbarossa2014}. On one hand, SBSs are connected with the cellular core via backhaul links, and on the other hand, SBSs are connected with mobile devices for service provisioning. Also, exploiting the real-time connection of SBSs with the core network assists SBSs to achieve computing-related functions such
as in-network caching using network-wide statistics. Here, both network providers and cloud operators must undertake effective cooperation towards optimizing the sharing of backhaul links, spectrum, and server utilization. Furthermore, the interplay between cloud and edge computing can support a broad diversity of IoT applications having different QoS requirements \cite{Sharma2017live}.

\subsection{Intelligent Edge-IoT Systems}
The enormous amount of data collected from IoT devices constitute an outstanding source of data-sets that can assist in training a machine and learning various features. This characteristic has enabled the possible utilization of ML techniques in a variety of computing aspects such as content caching, task offloading and device clustering to empower the decentralized operation of distributed IoT systems with the assistance of a central-cloud. In the following, we discuss important components of intelligent edge-IoT systems highlighted in Fig. \ref{fig:model}.

\subsubsection{Content Caching in Edge-devices}
The recent advances in cellular architectures such as H-CRANs enable network elements to have a comprehensive view on all the available resources. Also, the existence of SBSs in the vicinity of mobile users provides accurate spatio-temporal information about popular contents, and hence more accurate content caching and better usage of the limited storage capacity.

\subsubsection{Big Data Analytics with Edge Computing}
The proximity of edge-devices to mobile users ease the analysis and evaluation of big data collected from the computing hungry applications especially those equipped with signal processing functions through aggregation, filtering and pre-processing \cite{abbas2017}. Aided by the emerging ML techniques, big data can be harnessed to achieve the accurate prediction of popular contents, cache these contents in cellular SBSs, and provide timely feedback to the end-users without adding burdens on the backhaul resources \cite{zeydan2016, Sharma2017live}. However, the conventional complex data analysis algorithms suitable for the cloud should be simplified to be applicable for the resource- and computing-constrained edge-devices \cite{sahni2017}.

\subsubsection{Deep Learning for Edge Computing}
To utilize the benefits provided by both the cloud and edge computing platforms, decisions have to be intelligently made depending on the real-time training data, and learning-based approaches using both long-term and short-term statistics seem promising here to cope with the heterogeneous and dynamic nature of wireless IoT systems. For instance, offloading tasks that have stringent delay-sensitive requirements can be decided based on the instantaneous deterministic conditions. Whereas, decisions made for applications categorized with less delay-sensitive requirements could be more efficient when made depending on the stochastic long-term system performance \cite{mao2017}. To this end, several ML approaches (e.g., Neural Networks, Support Vector Machine, K-Means and Linear Regression \cite{MAHDAVINEJAD2017}) are available to perform intelligent decisions in several network operation tasks including load balancing, fault management and adaptive resource allocation.

\subsubsection{IoT Device Clustering}
Instead of sending all IoT data to the edge-server or central-cloud, IoT devices can be classified and clustered according to particular features such as functionality and geographic locations. For instance, home sensors can be processed by a central home controller, which can be equipped with ML capabilities to detect any unusual data (e.g., air pollution, noise and doors opened during unusual daily time) and forward these data to the homeowner or specialized authorities. Moreover, within a neighborhood, data collected from sensors that experience similar conditions such as location, environment and climate measurements can be further processed and verified by a neighborhood controller. This way, redundant data transmission from the thousands of devices can be avoided and hence reducing both communication and computing overheads. In addition, inter-cluster communication can be maintained by using different media such as wireless (e.g., WiFi or SBSs) and wired (e.g., fiber or ADSL).

\subsection{Flexible Infrastructure Management}
\label{reconfigurable}
Network management and control in the IoT era will face inevitable complexity due to the unprecedented density and heterogeneity of connected things. This necessitates the implementation of flexible management technologies to maintain adaptability, re-configurability and scalability.

\subsubsection{SDN}
The deployment of a layered fog architecture demands for flexible packet forwarding schemes over multiple fog layers on one hand, and from fog layers to the cloud layer on the other hand. The SDN technology helps to facilitate this process especially with the heterogeneity and mobility features of IoT devices (e.g., wearable devices). Moreover, fog nodes can be integrated with a particular cellular BS or shared by multiple BSs. Here, the SDN-based cellular core which adopts OpenFlow controllers and protocols can further ease the data forwarding process by facilitating control, mobility management, authentication, and network virtualization. Furthermore, when a device roams between different fog nodes, SDN helps in calculating the optimal profit regarding the end-to-end delay and cellular traffic, and triggers the Virtual Machine (VM) migration command according to the cost evaluation \cite{sun2016}.

\subsubsection{NFV}
By virtualizing different network functions on the shared hardware, NFV technology is capable of providing better control on the infrastructure, data sharing and network security. Moreover, it reduces the complexity of network administration and IoT system management \cite{lingeen2017}, and plays a significant role in system performance enhancement by scaling radio, computing, and infrastructure resources according to run-time needs \cite{barbarossa2014}. Some NFV example scenarios include Content Delivery Networks (CDNs) and the Platform as a Service (PaaS) model which allows users to use a particular software on shared machines.

\subsubsection{Flexible Radio}
The emerging cognitive radio technologies can enable cellular BSs to adapt their radio resources according to demand, allocate the best frequency resource, and decide which network tier is more suitable to serve a particular request (e.g., macro, pico, femto-cell).

\begin{figure*}[h]
\centerline{
\tcbox[sharp corners, boxsep=2mm, boxrule=0.2mm,
            colframe=black, colback=mblue]{
\includegraphics[scale=0.50]{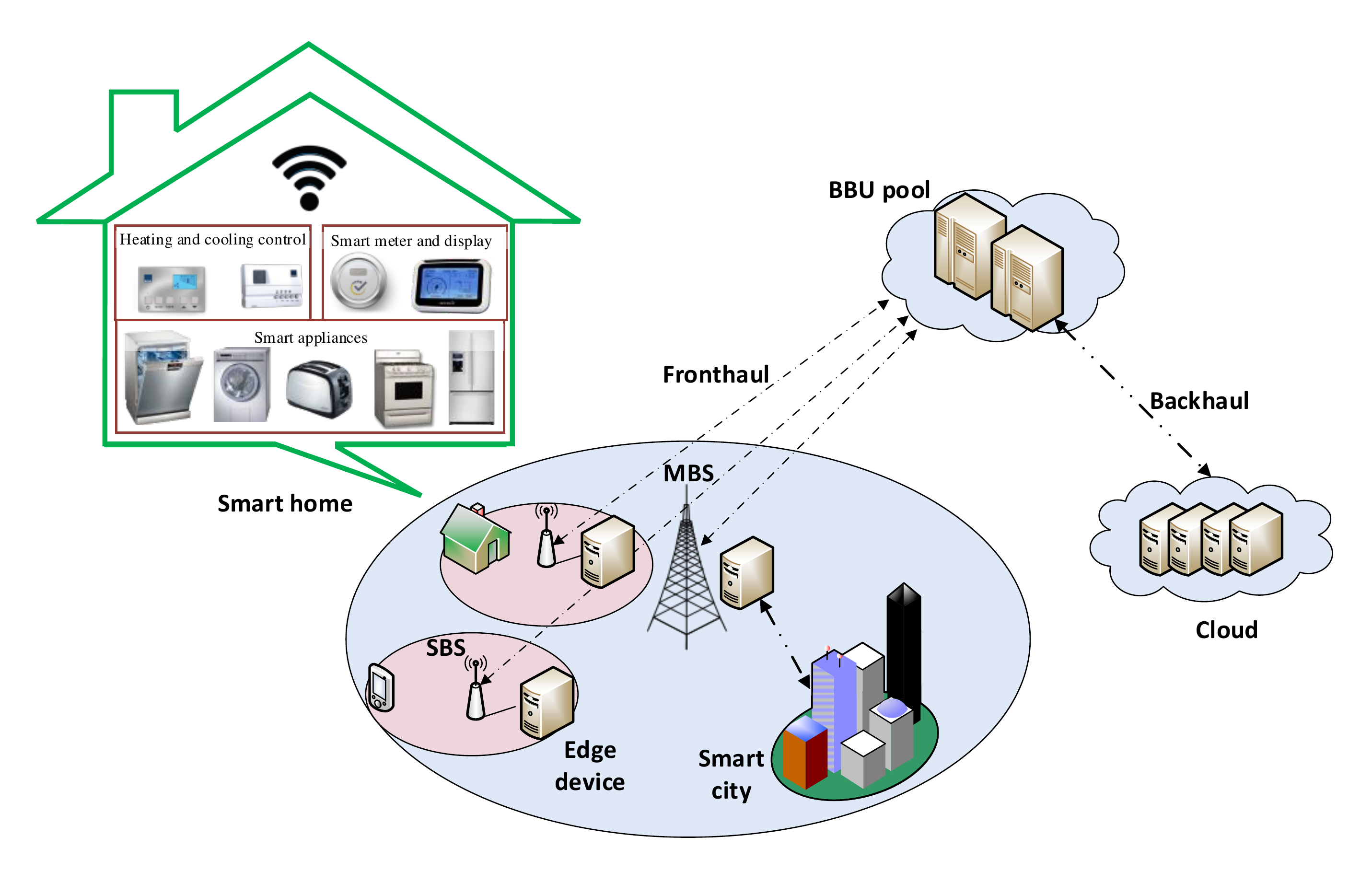}
}}
\caption{\small{Integrated cellular-computing system for IoT systems. The cellular part of the system is represented by the H-CRAN consisting of an MBS, multiple SBSs, and the central BBU pool, while the computing part comprises a hierarchical structure of a central-cloud and distributed edge-devices.}}\label{fig:model}
\end{figure*}

\section{Proposed Optimization Framework for Edge-IoT Systems}
Herein, we present an integrated cellular-computing architecture, which comprises of cellular elements such as SBSs, an MBS, a BBU pool, backhaul links with the computing nodes represented by edge-devices and a cloud as depicted in Fig. \ref{fig:model}. Also, the Key Performance Indicators (KPIs) of this architecture for different applications are highlighted in Table \ref{KPI}. It should be noted that the KPIs for one IoT application could be different from other applications. For example, energy is more critical for e-health and face recognition systems that rely on the energy stored in batteries compared to other systems that have continuous supply of energy such as homes and vehicles.

{\fontsize{12pt}{14.4pt}\setlength{\arrayrulewidth}{0.4mm}
\begin{table*}[h]
  \caption{\small{Key Performance Indicators (KPIs)}}\label{KPI}
\begin{center}
\begin{tabular}{>{\columncolor{dblue}}c|>{\columncolor{mblue}}c|>{\columncolor{dblue}}c|>{\columncolor{mblue}}c|>{\columncolor{dblue}}c}
\hline
    \backslashbox{\textbf{KPI}}{\textbf{IoT}} & e-Health & Face Recognition & Vehicular Communications & Home Sensors \\ \hline \hline

Throughput     & \checkmark & \checkmark  &  &  \\  \hline
Response Time  & \checkmark & \checkmark  & \checkmark &  \\ \hline
Energy & \checkmark & \checkmark  & &  \\  \hline
Security       & \checkmark & \checkmark  & \checkmark & \checkmark \\

     \end{tabular}
  \end{center}
\end{table*}
}


\begin{figure*}[h]
\centerline{
\tcbox[sharp corners, boxsep=2mm, boxrule=0.2mm,
            colframe=black, colback=mblue]{
\includegraphics[scale=0.55]{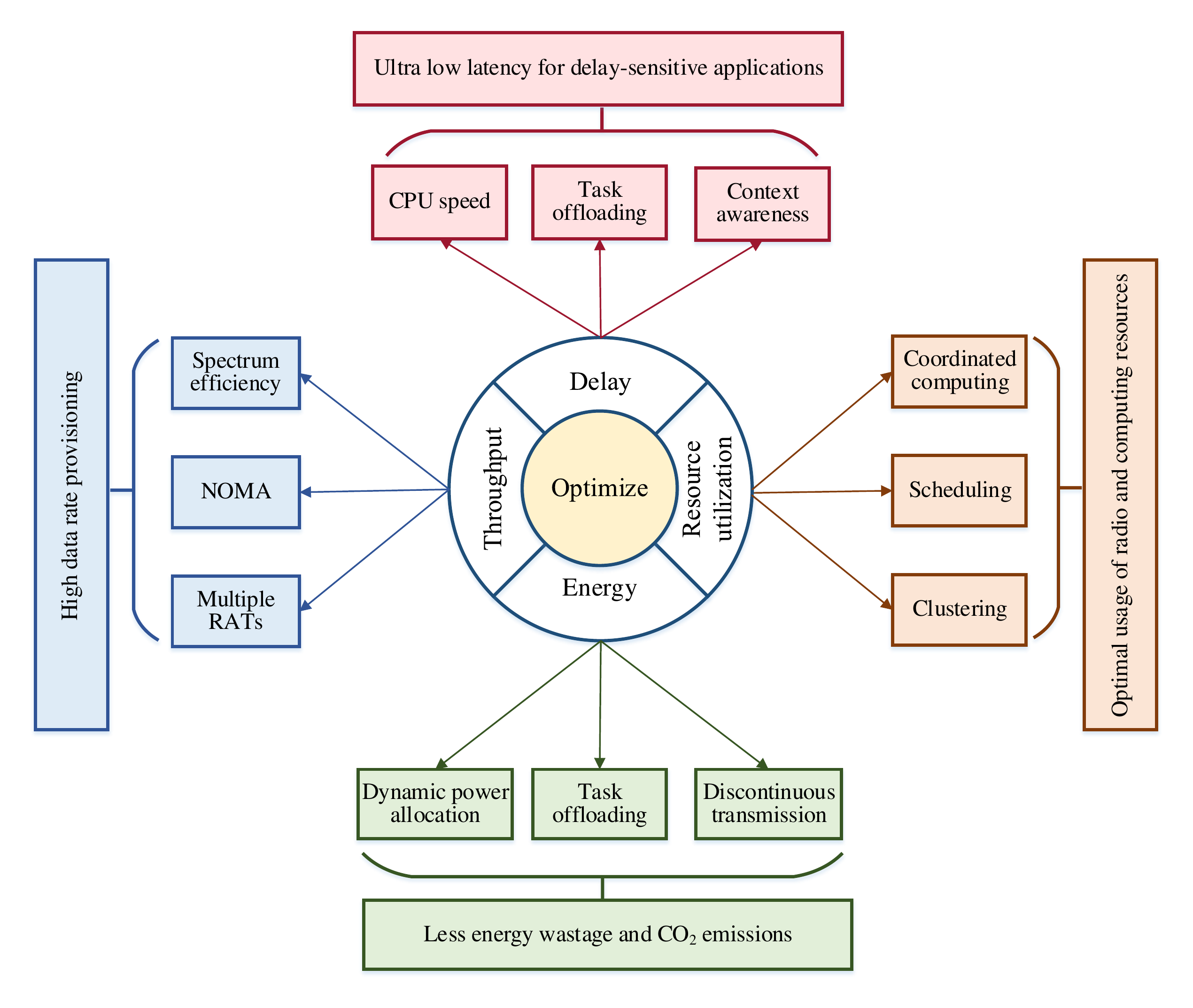}
}}
\caption{\small{Proposed optimization framework for distributed IoT systems. The KPIs considered in the framework are delay, resource utilization, energy and throughput. Potential solutions are identified for each KPI to effectively reach the desired optimization goals.}}
\label{fig:resource}
\end{figure*}

Figure \ref{fig:resource} presents the proposed optimization framework which aims to achieve the following four objectives: (i) ultra-low latency for delay-sensitive applications, (ii) optimal usage of radio and computing resources, (iii) less energy wastage and CO$_2$ emissions, and (iv) high data-rate provisioning. For the first objective, delay is a crucial performance metric and can be minimized by optimizing CPU speed, task offloading and context awareness. Whereas, for the second objective, resource utilization can be enhanced via coordinated computing, effective device clustering, and intelligent learning techniques. Similarly, for the third objective, energy efficiency is of the main concern in IoT systems and can be optimized through dynamic power allocation, smart task offloading decisions, and discontinuous transmission. Regarding the fourth objective, several enablers such as small cells, NOMA and multiple RATs can enhance throughput in edge-IoT systems.

In the following, we briefly describe some promising approaches to improve four performance metrics highlighted in Fig. \ref{fig:resource}.

\begin{enumerate}
\item \textbf{Delay}: Dealing with delay-sensitive tasks is more critical than other tasks because delay consequences can be severe for some applications such as autonomous driving. Thus, task completion deadline and the amount of CPU cycles required for each particular context-awareness device must be considered before a decision is made on whether to perform on-device processing, or to offload tasks to the edge-server/central-cloud. Moreover, the arrival rate of incoming tasks must be incorporated in such decisions to avoid overloading server buffers that negatively impact the delay experienced by IoT devices. In other words, long-term optimization approaches such as Markov decision process can be useful to optimize the overall system performance.
\item \textbf{Resource utilization}: To take full advantage of edge computing, a high-level task coordination among the all computing parties (on-device, edge, and cloud processors) is crucial and towards achieving this goal, novel central, distributed and hybrid scheduling schemes can be investigated. Also, radio resource scheduling via an SDN- and NFV-based BBU pool helps to boost the computing capabilities by optimally allocating frequency resources to IoT devices taking the bandwidth and interference constraints into account. Furthermore, the formation of the volatile on-demand fog nodes can support edge computing in saving resources and reducing the response time. Device clustering is another form of NFV-based resource saving in edge computing whereby one VM can be shared by multiple devices to maximize the usage of computing resources. Moreover, learning-based approaches seem promising to dynamically adapt to the heterogeneity of IoT demands.
 \item \textbf{Energy}: As one of the major constraints in future communication and computing systems, energy consumption must be minimized not only to prolong the lifetime of on-device batteries but also to reduce the massive amounts of CO$_2$ emissions. Some of the efficient approaches to address this issue include efficient power allocation schemes, short-range transmission through SBSs, optimization of CPU cycles, task offloading, discontinuous transmission and hardware sleeping schemes.
  \item \textbf{Throughput}: The computing tasks processed on the powerful cloud and edge processors must be complemented by high data-rate provisioning to avoid undesired latencies. To this goal, exploiting the dense deployment of SBSs can enable the reuse of frequency resources over small areas, and hence increase the per-user data rate. Other technologies that can enhance the spectral efficiency include NOMA and massive MIMO. Also, integrating many RATs such as 5G, LTE, WiFi and optical fibers can diversify the supply of frequency resources.
\end{enumerate}

\section{Use Case Study}
Herein, we present a use case study to demonstrate the significance of an ML-assisted solution for the virtual clustering of distributed IoT devices. To avoid the congestion of backhaul-links, IoT devices can be virtually clustered at the edge forming different virtual groups. In such a scenario, Reinforcement Learning (RL)-based data classification and aggregation can help not only in saving radio and computing resources but also in supporting a scalable computing paradigm that can efficiently adapt to sudden changes in data volumes and traffic variations. Moreover, processing data on the shared machines facilitates the local modification of cloud mechanisms by edge-devices. Therefore, an ML-assisted solution is employed here for the considered virtual device clustering problem.

To initiate a cluster of distributed IoT devices, the edge-device/aggregator assigns a cluster-head that has more powerful computing capabilities than other cluster members, and becomes in charge of managing intra-cluster communications and cluster-edge communications. The formation of clusters (small fogs) is performed by the edge-devices based on physical proximity, function or correlation coefficients among the clustered devices. For instance, smart home sensors in a particular neighborhood that measure the outside temperature have a common function and almost similar data; thus, it is more efficient in regard with computing, spectral and energy efficiencies to forward only abstracted meaningful information to the central-cloud. The clustering process can be successfully achieved by using intelligent and powerful ML capabilities that analyze the daily behavior of sensors, extract the relationships between different data, and cache necessary data at the edge-device or central-cloud for future referencing.
\begin{figure}[h]
    \begin{center}
    \tcbox[sharp corners, boxsep=2mm, boxrule=0.2mm,
            colframe=black, colback=white]{
    \begin{subfigure}[b]{0.29\textwidth}
        \includegraphics[width=\textwidth]{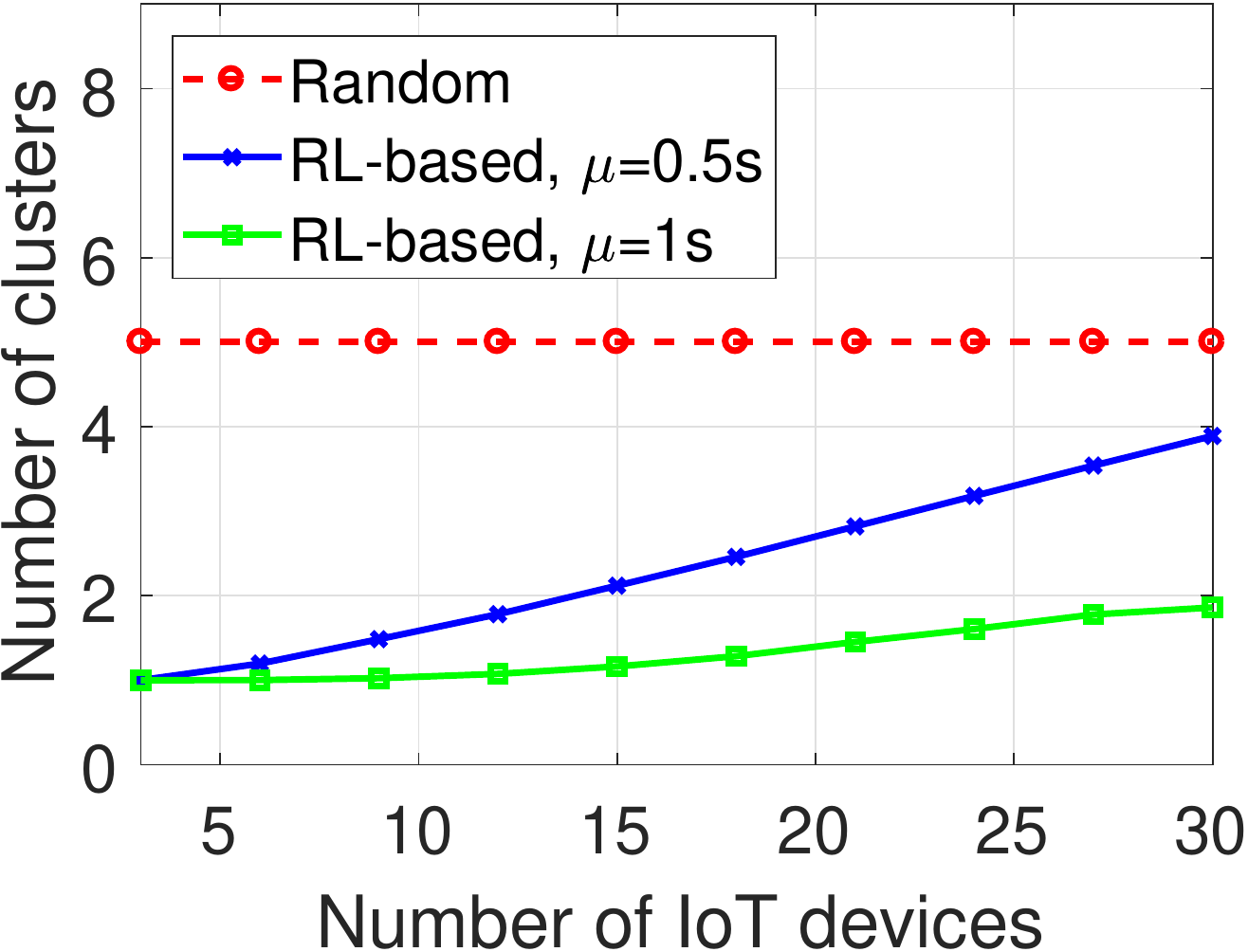}
        \caption{Number of clusters required in the proposed RL-based and random schemes.}
   \end{subfigure}
~\hspace{0.3cm}
    \begin{subfigure}[b]{0.29\textwidth}
        \includegraphics[width=\textwidth]{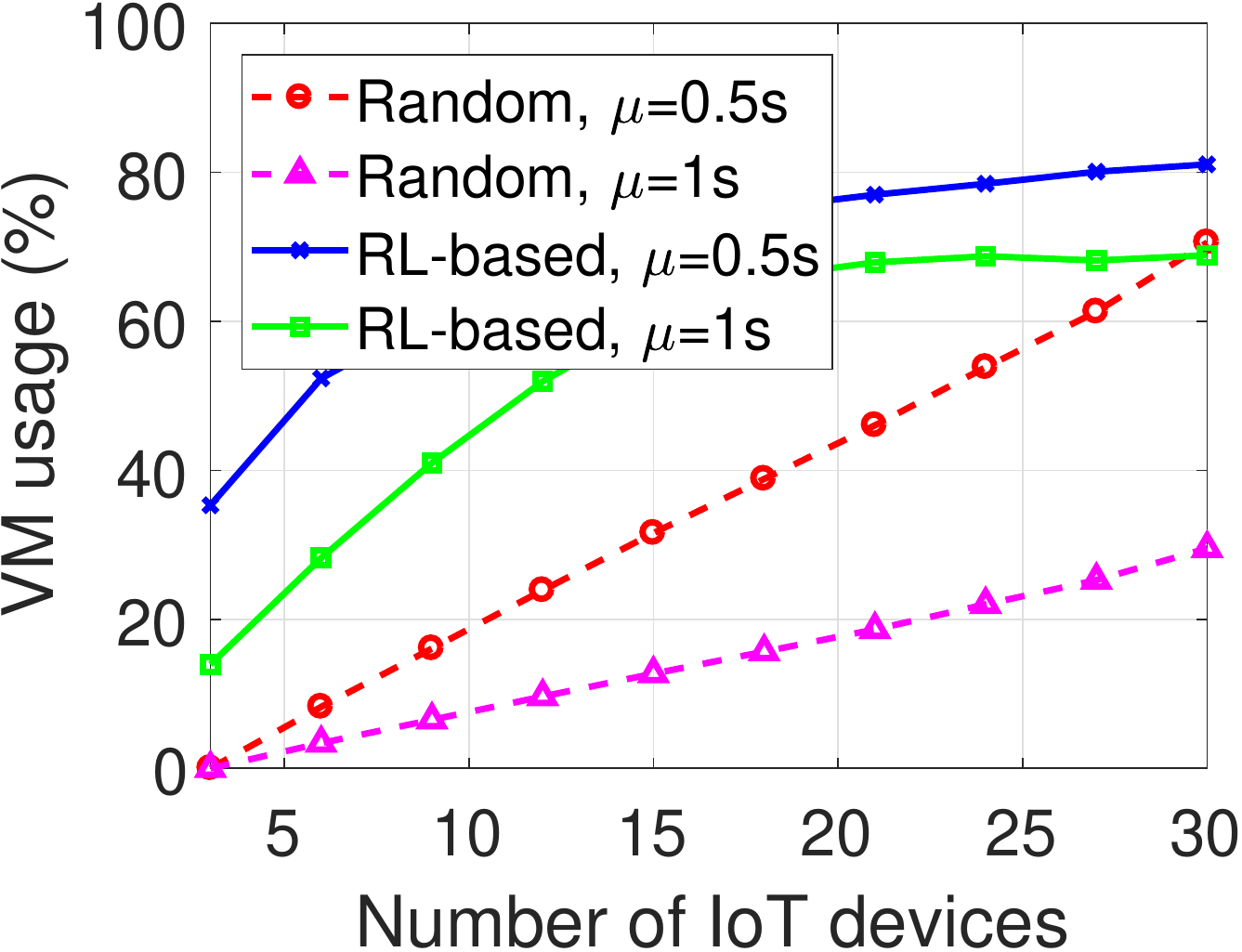}
        \caption{Usage of computing resources (VM) increases with delay requirements.}
   \end{subfigure}
~\hspace{0.3cm}
    \begin{subfigure}[b]{0.29\textwidth}
        \includegraphics[width=\textwidth]{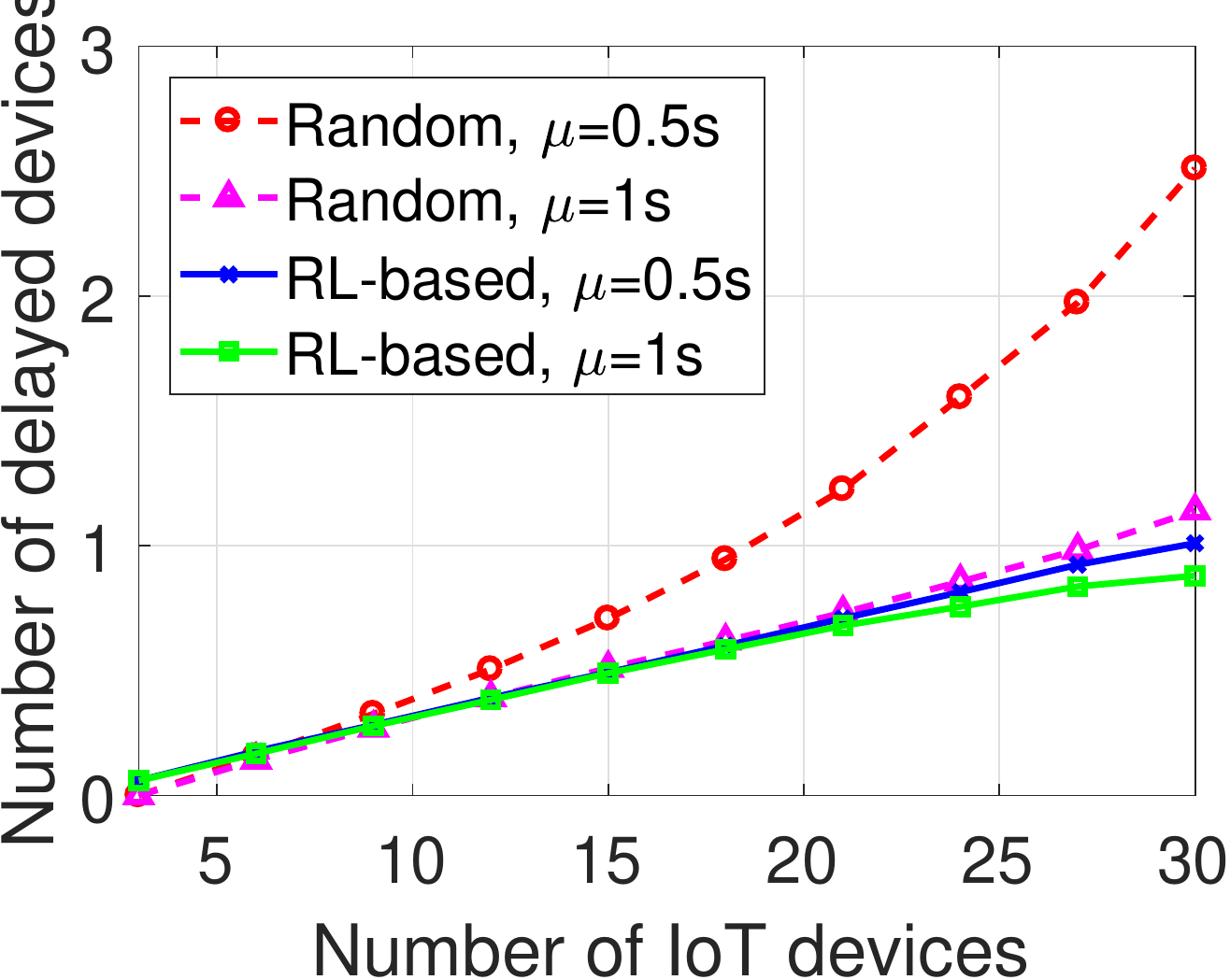}
        \caption{Average number of delayed IoT devices in the system.}
   \end{subfigure}}
    \caption{\small{RL-based versus random device clustering in edge-IoT systems (The number of available VMs is $5$, the discount factor and learning rate for the RL-based approach are $0.9$ and $0.1$, respectively).}} \label{casestudy}
    \end{center}
\end{figure}

By exploiting the possibility of clustering numerous IoT devices under one cluster head, the amount of radio transmissions from IoT devices to the edge-device can be significantly decreased, thus reducing the overall energy consumption. Clusters in the system under consideration are assumed to be formed according to the physical proximity where a cluster head is in charge of maintaining the communication between cluster members and the edge-device that assigns one VM to each cluster. Each IoT device is assumed to have a particular packet size and the completion deadline to accomplish the desired task. The packet size is uniformly distributed between $500$ kb and $4$ Mb whereas the task completion deadline is assumed to differ between two groups; the first group has a deadline range between $100-900$ ms (i.e., mean delay $\mu=0.5ms$), while the second group has the deadline range of $500-1500$ ms (i.e., mean delay $\mu=1 s$). The aim of this study is to investigate the optimal number of IoT devices per cluster (VM) such that less VMs are to be used while satisfying the deadline requirement of each IoT device. Here, an RL technique, namely Q-learning, is utilized to address the aforementioned problem. Two actions are considered during Q-learning process, namely, increment and decrement such that the reward for incrementing the number of members in a cluster is $+5$ and for decrementing the number of members is $-1$. Whereas, having a delayed IoT device turns the reward for incrementing and decrementing the number of members to $-10$ and $5$, respectively.

Figure \ref{casestudy} depicts the system performance under different delay constraints. In the random scheme, IoT devices are randomly distributed among the $5$ VMs present in the system. It can be observed that RL-based technique that aims to cluster IoT devices on shared VMs taking into account the delay requirement of each device provides better performance in reducing the amount of wasted resources which is represented by the smaller amount of allocated VMs (clusters) as depicted in Fig. \ref{casestudy} (a). Moreover, the RL-based approach can increase the percentage of VM utilization, thus leading to more efficient resource utilization as shown in Fig. \ref{casestudy}(b). Also, it can be noted that having more stringent delay requirement demands for more computing resources (VMs), and subsequently increases the number of delayed devices as observed in Fig. \ref{casestudy}(c).

\section{Conclusions}
Edge computing along with the central-cloud constitutes a powerful computing paradigm to enable the practical realization of distributed IoT systems. However, there still exist several issues from both the communication and computing perspectives, and various technologies such as cooperative resource management, ML, context-aware computing, and flexible infrastructure utilization using NFV and SDN are being emerged in this direction. In this regard, this article provided a comprehensive view on the existing research issues and emerging edge-computing technologies, and proposed a novel framework for the optimization of key performance metrics in edge-IoT systems.

The upcoming IoT era calls for serious efforts in the direction of ML, SDN, and NFV to establish self-organized and self-resilient computing systems and to cope with the heterogeneity of IoT services. Furthermore, since communication and computing are inherently integrated in IoT systems, computing system optimization must take into account of the constraints imposed by the limitations of cellular networks no matter whether these constraints are related to insufficient resources, time-varying wireless links, or other issues such as energy and cost. In other words, joint communication-computing systems with learning features, and flexible infrastructure and resource management need to coexist simultaneously with the distributed IoT systems.
	


\end{document}